\documentclass[prl,twocolumn,amsmath,amssymb]{revtex4-1}

\usepackage{graphicx}
\usepackage{color}

\begin{document}

\title{Schottky-to-Ohmic Crossover in Carbon Nanotube Transistor Contacts}

\author{V. Perebeinos}
\affiliation{IBM T.J. Watson Research Center, Yorktown Heights, NY 10598, USA}
\author{J. Tersoff}
\affiliation{IBM T.J. Watson Research Center, Yorktown Heights, NY 10598, USA}
\author{W. Haensch}
\affiliation{IBM T.J. Watson Research Center, Yorktown Heights, NY 10598, USA}

\date{\today}


\begin{abstract}

For carbon nanotube transistors, as for graphene,
the electrical contacts are a key factor limiting device performance.
We calculate the device characteristics as a function of
nanotube diameter and metal workfunction.
Although the on-state current varies continuously,
the transfer characteristics reveal a relatively abrupt crossover
from Schottky to ohmic contacts.
We find that typical high-performance devices fall
surprisingly close to the crossover.
Surprisingly, tunneling plays an important role even in this regime,
so that current fails to saturate with gate voltage as was expected
due to ``source exhaustion''.

\end{abstract}

\maketitle

Since the earliest studies of carbon nanotube field-effect transistors (CNT FETs),
the metal contacts have been a key factor limiting device performance.
Early contacts were invariably dominated by Schottky barriers
\cite{Martel01,Heinze_Schottky,Joerg}.
A major breakthrough came with the fabrication of
robust low-resistance contacts \cite{Javey03},
although so far only for p-type contacts.
Correspondingly, models of CNT contacts generally focus
on one of two simple regimes.
For Schottky contacts, the focus is exclusively on transmission through the barrier,
with the gate serving to thin the barrier and increase tunneling.
For ohmic contacts, incidental local barriers can still play some role;
but for many purposes these can be neglected,
especially for thin gate oxides \cite{Leonard_APL}.
Then for ballistic devices,
the on-state current $I_{on}$
is expected to be limited by ``source exhaustion'':
the maximum possible current is set by the contact doping,
i.e.\  the number of carriers provided by charge transfer from the metal,
times their velocity \cite{starvation_Guo,starvation_Fischetti,starvation_Wong}.

This simple dichotomy has proven adequate for general discussions,
but it has never been directly confirmed by experimental measurements.
Even in nominally ohmic CNT FETs it is difficult to distinguish source exhaustion
from other effects that could cause $I_{on}$ to saturate with increasing gate voltage.
More importantly, the height of any Schottky barrier is expected to vary
in a simple way with metal workfunction and CNT bandgap.
Therefore one might expect a strong dependence of $I_{on}$
on these factors, up to the point at which the barrier vanishes,
and a much weaker dependence in the ohmic regime.
However, the only experiment to directly address this
reported a strikingly continuous dependence spanning the entire range
from good devices to high-resistance contacts \cite{Zhihong}.

Here we calculate the behavior of CNT-FETs
using a device model that is applicable across both regimes.
Consistent with Ref.~\cite{Zhihong},
we find that when one simply examines on-state current $I_{on}$,
there is only a smooth variation with metal workfunction
and CNT bandgap, no clear transition between regimes.
However a clear transition is present in other performance measures,
with a qualitative change in the shape of the transfer characteristics.
Comparison with experiment indicates that
typical high-performance devices operate in a regime
surprisingly close to the Schottky-ohmic crossover.
This suggests the possibility of further improvements
in device performance via workfunction engineering.

We find that device characteristics in the ohmic regime
are rather different than expected.
In particular, simple source exhaustion is not observed in our calculations.
This is fortunate, because current saturation with gate voltage
is undesirable for transistors.
Instead, the current continues to rise with increasing gate voltage.
The reason is that, while the doping in the CNT is limited,
there are plenty of carriers in the metal at energies within the CNT bandgap.
These metal states can tunnel to the channel via evanescent states
in the CNT \textit{underneath the metal contact}.
This tunneling increases continuously with gate voltage,
and is particularly large for the thin gate oxides
used in advanced high-performance devices.

Our computational method is an extension of semiclassical device modeling
to include tunneling, as well as electronic coupling between the
CNT and the metal in the usual side-contact configuration. All energies are measured relative to ground, i.e.\  to the source Fermi level. For energies outside the bandgap,
the distribution function $f^{r}_i(x,E)$ for right-moving carrier resolved by energy $E$ and band index $i$ obeys:
\begin{eqnarray}
0&=&-\frac{df_{i}^{r}(E,x)}{dx}+\frac{f_{i}^{r}(E,x)-f^{0}(E-E_{F}^{M}(x))}{v_{i}(E,x)\tau_{M}(E,x)}
\nonumber \\
&&+\frac{f_{i}^{r}(E,x)-f^{0}(E-E_{loc}(x))}{v_{i}(E,x)\tau_{scat}(E,x)}
\label{eqdfdt}
\end{eqnarray}
where $v_i(E,x)$ is the band velocity and $f^0$ is the Fermi-Dirac distribution.
The second term describes an electrical coupling to the metal, which provides a source of current into and out of the nanotube \cite{Datta}.
For left-moving carriers $f^{l}_i(E,x)$, the sign of $v_i(E,x)$ in Eq.~(\ref{eqdfdt}) is opposite.
The third term allows thermalization of the carriers toward an equilibrium distribution within a relaxation-time approximation,
with the local quasi-Fermi level $E_{loc}(x)$ found self-consistently.
Further details of the model are given in Ref.~\cite{suppl}.
We can express current density $ j_i(E,x)$ resolved by energy and band index, and total current $I$,  as:
\begin{eqnarray}
j_i(E,x)&=&\frac{4e}{h}\left(f_i^r(E,x)-f_i^l(E,x)\right)
\nonumber \\
I(x)&=&\sum_{i}\int_{-\infty}^{\infty} j_i(E,x) dE
\label{eqcurr}
\end{eqnarray}
At a given energy, if two regions of propagating states
(including metal states) are separated by a region with
no states at that energy, we use the tunneling probability
to set boundary conditions for transmission and reflection
at the classical turning point.  (Inelastic tunneling is not considered.)
Carrier transfer at each point along the metal-CNT side contact
is treated assuming an energy-independent transfer rate.
In this way, we can consistently describe both ballistic and diffusive transport,
with Schottky or ohmic contacts, including ambipolar devices.

\begin{figure}[b]
\includegraphics[width=3.4in]{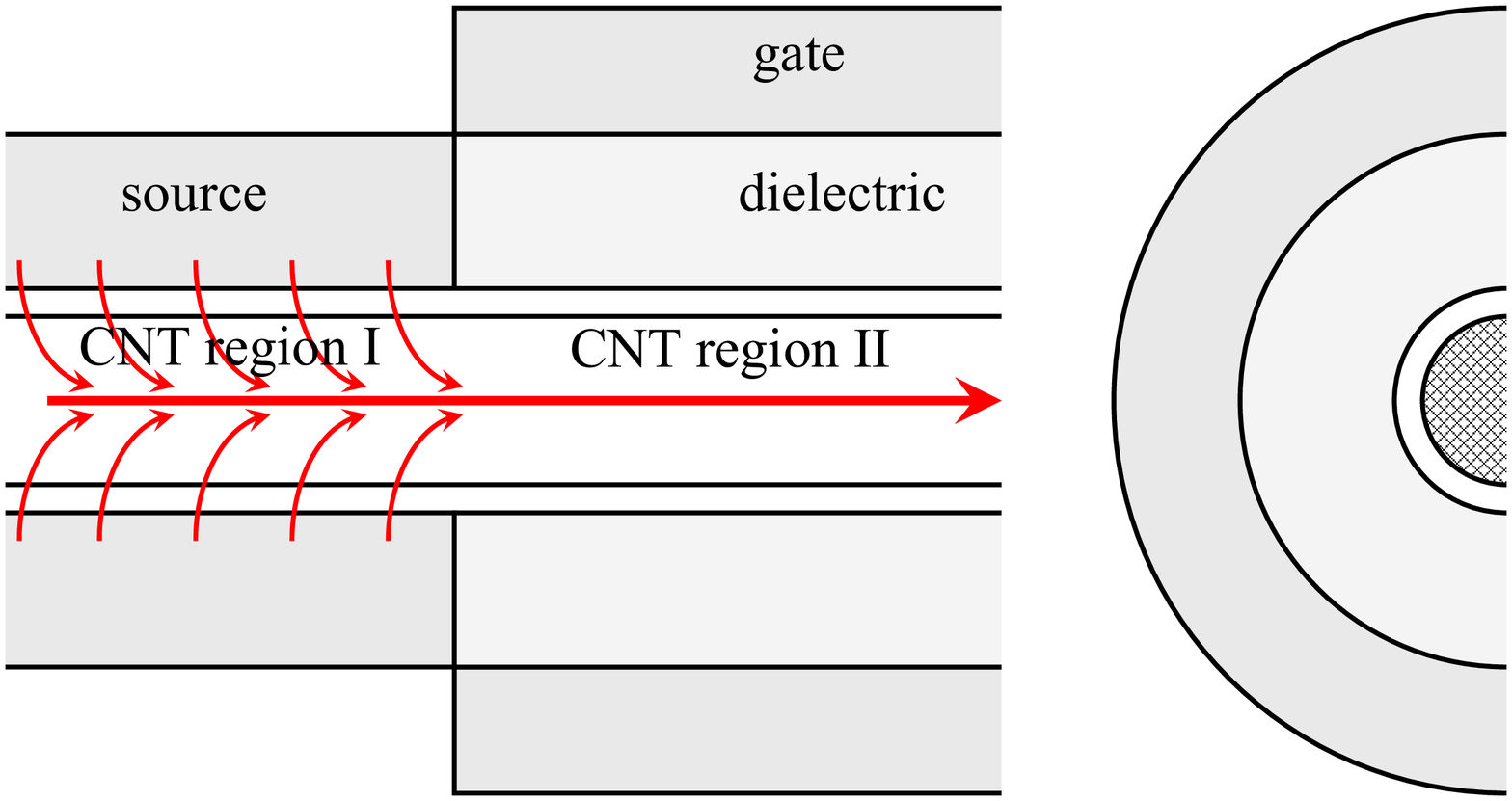}
\caption{ Gate-all-around device geometry used here. Left is cross section of contact region, right in view down axis.  The red arrows schematically illustrate the current paths from metal to the channel.}
\label{FigSchematic}
\end{figure}

We focus on nearly ballistic devices in the ideal cylindrical geometry, in which both
the contact and gate wrap around the CNT,
as shown in Fig.~\ref{FigSchematic}.
That figure also illustrates schematically the current
pathway from the channel to the metal.
In a Schottky contact, the source Fermi level falls
in the CNT bandgap, so the states in CNT region I are evanescent
(regardless of $V_g$),
but they still provide a path for tunneling between the metal
and the channel.

The carrier distribution function $f^{r(l)}(E,x)$
is calculated self-consistently,
with the metal gate and contacts providing electrostatic boundary conditions.
The CNT multi-band electronic structure is described by a one-parameter
(pi-only) tight-binding model.
Device parameters are chosen to facilitate comparison with experiment \cite{Zhihong},
see \cite{params-Yan} for specific values.
Following Ref.~\cite{Zhihong} we define the on-state current $I_{on}$ by
overdrive $V_g-V_t = -0.5$ V. The drain voltage $V_d = -0.5$ V,
which is usually enough to saturate current in the regime of greatest interest.

\begin{figure}[b]
\includegraphics[width=3.4in]{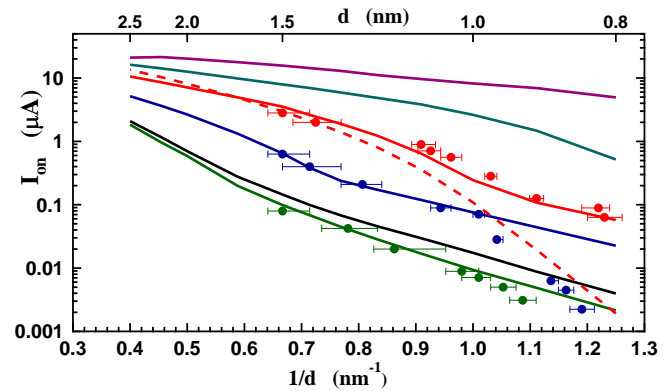}
\caption{
On-state current $I_{on}$ vs CNT inverse diameter
(proportional to bandgap), for different metal workfunctions from top to bottom:
$\Delta W= W_m-W_{c}=$
0.8, 0.5, 0.35, 0.23, 0.0, -0.1 eV.
The gate dielectric has thickness $t_{ox}=20$ nm,
and $\epsilon=3.9$ as for SiO$_2$.
The symbols and error bars show experimental data from Ref.~\protect{\cite{Zhihong}}
for Pd, Ti and Al metals, shown from top to bottom in
red, cyan, and green circles respectively.
(We show only data for the range $1/d<1.3$ nm$^{-1}$ recommended by those authors
as relatively reliable.)
The dashed line shows the current in the source-exhaustion limit,
$I_{se}$ from Eq.~(\protect{\ref{EqImax}}),
for $\Delta W=0.35$ eV, based on the self-consistent
non-equilibrium carrier distribution.
}
\label{FigZhihong}
\end{figure}

Figure~\ref{FigZhihong} shows $I_{on}$ as a function of CNT diameter,
for metals with a range of workfunctions.
The qualitative trends are well understood
from simple band-alignment arguments.
Relative to the vacuum level, the metal Fermi level is at $-W_m$
and the CNT valence edge is at $-W_c-E_g/2$,
where $W_m$ and $W_c$ are the workfunctions of the metal
and a metallic CNT respectively, and $E_g$ is the CNT bandgap,
which scales with diameter as $E_g \propto 1/d$.
Thus for p-type contacts the Schottky barrier height is $E_g/2 + W_c - W_m$.
Ohmic p-type contacts are obtained by using large-workfunction metals,
especially Pd, in combination with small-bandgap
(i.e.\ large-diameter) CNTs \cite{Javey03},
so that the metal Fermi level falls near or below
the CNT valence band edge.

Experimentally, it is prohibitively difficult to measure
the actual CNT diameter in a statistic number of working devices.
Nevertheless one paper has reported the variation in $I_{on}$
with diameter, for contacts made with several different metals \cite{Zhihong}.
Those authors note that their CNT diameter values are not entirely reliable,
because they are inferred only indirectly,
using a statistical analysis with a strong auxiliary assumption
that variations in $I_{on}$ for a given metal
are due primarily to the CNT diameter.
Even assuming that is correct,
the statistics are only reliable for diameters in the middle of range sampled.
Nevertheless, in the absence of other data,
that seminal work provides a natural starting point for comparison,
and those data are included in Fig.~\ref{FigZhihong}.

We find a striking agreement between theory and experiment
over two orders of magnitude in $I_{on}$ in Fig.~\ref{FigZhihong},
if we take the workfunction
for each metal as a fitting parameter.
We actually use only the workfunction difference $\Delta W= W_m-W_c$
between metal and CNT midgap.
In agreement with Ref.~\cite{Zhihong} we find that the workfunctions
inferred in Figure~\ref{FigZhihong}
vary by less than expected from literature values of the workfunctions.
Most importantly, the fitted value $\Delta W=0.35$ eV for Pd contacts
is substantially smaller than the expected range
$\Delta W\approx 0.9 \pm 0.2$ eV.
This difference might simply reflect the fact that literature values
of workfunction are for ultra-clean surfaces in values,
while the actual devices
have had prolonged exposure to air, and both metal and CNT
may have picked up other impurities during the device processing.
Or more fundamental electronic-structure effects
may play a role \cite{Giovannetti,Khomyakov}.
Figure~\ref{FigZhihong} suggests the possibility of
further improvements in device performance via workfunction engineering.

Ti and Al are well known to give Schottky contacts,
and the behavior here is consistent with previous modeling.
Narrow CNTs have larger bandgaps, giving larger barrier heights.
But the current decreases more slowly than for thermally activated transport,
because it is dominated by tunneling.
We note that in this regime,
the detailed shape of the contacts as well as the workfunction
can substantially influence the results
\cite{Heinze_Schottky, Heinze_scaling}.

\begin{figure}[b]
\includegraphics[width=3.4in]{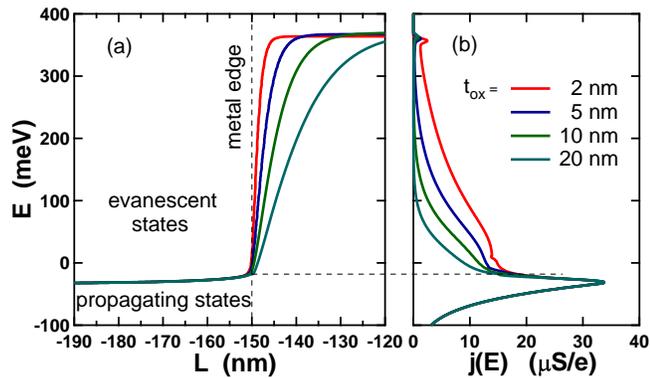}
\caption{
Band bending and current distribution for a d=1.3 nm CNT and metal
$\Delta W=0.35$ eV at overdrive $V_g-V_t=-0.5$ V
and $V_d=-0.5$ V, for gate-oxide thicknesses
$t_{ox}=2$, 5, 10, 20 nm from top to bottom respectively.
a) Valence band edge vs position.
The zero of energy is the source Fermi level.
The vertical dashed line show position of the
source metal edge at $L=-150$ nm.
b)  Energy-resolved current density, Eq.~(\protect{\ref{eqcurr}}),
in the middle of the channel.
Horizontal dashed line
separates energies where CNT states are
propagating vs evanescent in the source contact.
The current contribution from the propagating states in (b)
is $I_{se}\approx 1.3$ $\mu$A, nearly independent of $t_{ox}$.
The total current is
$I\approx 4.0$, 3.0, 2.4, and 2.0 $\mu$A for
$t_{ox}=2$, 5, 10, 20 nm respectively.
}
\label{FigBending}
\end{figure}

\begin{figure}[b]
\includegraphics[width=3.4in]{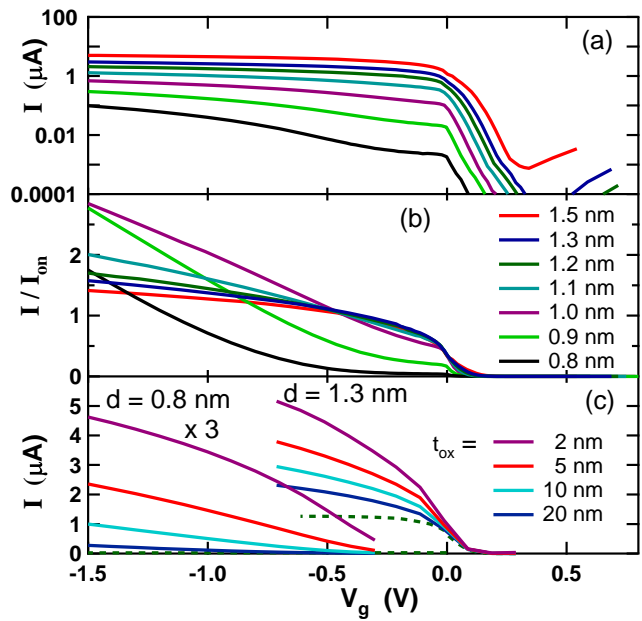}
\caption{
(a) Transfer curves $I$ vs $V_g$  for different CNT diameters,
for $V_d=-0.5$ V, $t_{ox}=20$ nm, and $\Delta W=0.35$ eV.
(b) Same as (a) but on a linear scale, with each curve
normalized to $I_{on}$ at $V_g=V_t-0.5$ V.
($I_{on} =$ 3.5, 2.6, 1.9, 1.2, 0.64, 0.24, 0.11, and 0.057 $\mu$A
for $d =$  1.5, 1.4, 1.3, 1.2, 1.1, 1.0, 0.9, and 0.8 nm respectively.)
(c) $I$ vs $V_g$  for CNT diameters $d$=0.8 nm (left) and $d$=1.3 nm (right),
with $\Delta W$=0.35 eV.
Each diameter is shown for $t_{ox} =$ 2, 5, 10, 20 nm from top to bottom.
Note the multiplicative factor for $d=0.8$ nm CNT in (c).
}
\label{FigIvg}
\end{figure}

The maximum possible current
is expected to be governed by ``source exhaustion'' and
``source starvation'' \cite{starvation_Guo,starvation_Fischetti,starvation_Wong}.
This applies for ballistic devices with ideal ohmic contacts,
when the drain current $V_d$ is large enough to reach saturation.
In this limit, all the carriers in the CNT at the source end
that are moving in the direction toward the channel
are transmitted to the drain with probability 1.
This carrier density depends on the self-consistent electrostatics
of the CNT under the metal, with the actual nonequilibrium population of
outgoing carriers.
The corresponding current is
\begin{eqnarray}
I_{se}&=& (4e/h)k_BT\ln{\left[
1+\exp{\left(  E_{Fs}/k_BT   \right)}
\right]}
\label{EqImax}
\end{eqnarray}
where $E_{Fs}$ is the Fermi level in the CNT under the source metal contact,
relative to the valence bandedge.

The calculated current $I_{se}$ in this source-exhaustion limit
is shown as a dashed line in Fig.~\ref{FigZhihong} for
the case of $\Delta W=0.35$ eV (as for Pd contacts).
The actual current can be less even for ballistic devices,
because of additional electrostatic barriers \cite{Leonard_APL}.
We find that this is an important effect for very large workfunctions,
but for $\Delta W=0.35$ it is significant only for the largest diameters.
In general, $I_{se}$ gives a good semi-quantitative description
for the larger-diameter CNTs.

For narrower tubes, the doping in the tube and $I_{se}$ become
exponentially small as the Fermi level falls deep in the bandgap.
The current however decreases more slowly,
as it is dominated by tunneling to the metal,
which represents a separate additional transport mechanism.
From this perspective, it seems something of a coincidence that
there is not a more striking change of slope in Ion vs 1/d
when crossing between Schottky and ohmic regimes,
resulting from two quite different mechanisms giving a similar slope
in $I_{on}$ vs diameter.

Surprisingly, we find that $I_{on} > I_{se}$ for all devices
in the diameter range of the experimental data.
This indicates that tunneling is an important contribution
even for nominally ohmic devices.
To understand this, in Fig.~\ref{FigBending}
we show the energy spectrum of the transmitted current.
Here we focus on the case $\Delta W$=0.35 eV and $d$=1.3 nm,
comparable to the best experimental devices.
There is a large energy range where the source has
states in the metal but not in the CNT.
Carriers can tunnel from the metal to the channel
via the evanescent modes of the CNT underneath the metal,
as illustrated schematically in Fig.~\ref{FigSchematic}
(CNT region I),
and through the external barrier in CNT region II.
As the gate oxide gets thinner the barrier
becomes correspondingly thinner in Fig.~\ref{FigBending}(a),
and so the tunneling current becomes increasingly important in Fig.~\ref{FigBending}(b).
In contrast, the current contribution from propagating states
is seen to be virtually independent of the gate oxide thickness.

Even for the 20 nm oxide, tunneling gives a 50\% increase
in $I_{on}$ relative to the expected current $I_{se}$
from propagating states.
More advanced devices now use HfO$_2$,
which has a larger dielectric constant than SiO$_2$,
and which also allows fabrication of gate oxides as thin as 3 nm.
Even thinner gate oxides can be made using Si oxynitride.
We find that in the ballistic devices studied here,
the dielectric constant has little impact,
but the geometrical thickness is crucial.
For the thinnest oxides, we find that the majority of the current
comes from tunneling, even for the best devices.
The quantitative values obtained here apply only to the specific geometry
of Fig.~\ref{FigSchematic}, but the trends should apply very generally.

In Fig.~\ref{FigIvg} we show calculated transfer characteristics $I$-$V_g$
for Pd contacts in devices with different CNT diameters.
On a log scale, Fig.~\ref{FigIvg}(a), the characteristics
all appear rather similar.
The most obvious difference is an overall reduction in current
at smaller CNT diameters.
We also note the appearance of ambipolar behavior with increasing diameter.
(Even larger diameters would give higher current but lower on/off ratio,
making such CNTs unsuitable for FETs.)

Figure~\ref{FigIvg}(b) shows the same results
plotted on a linear scale, after normalizing each curve by $I_{on}$.
For the larger-diameter CNTs, we see that the curves
are still rather similar aside from the overall scaling by $I_{on}$.
There is a relative sharp step in the current
at threshold ($V_g \approx 0.11$ V).
The current continues to rise with increasing $V_g$,
and this non-saturation is proportionally greater for the
CNTs that are closer to the ohmic-to-Schottky crossover.
However for the narrower CNTs the curves have qualitatively different shape,
with no visible step in current, only a smooth increase.

In Fig.~\ref{FigIvg}(c) we show calculated transfer characteristics
for different oxide thicknesses, focusing on the case
$\Delta W$=0.35 eV and $d$=1.3 nm as in Fig.~\ref{FigBending}.
The dashed line shows the result for a 20 nm oxide,
if we artificially suppress all tunneling.
Results for other thicknesses are nearly the same
when tunneling is suppressed, and all show
the expected saturation corresponding to $I_{se}$.
Such saturation is highly undesirable, since current technology
is based on current increasing smoothly with $V_g$.
In contrast the full calculation shows no saturation with $V_g$,
and increasingly high currents due to tunneling for
thinner oxides and larger overdrives.
Ultra-thin oxides are considered desirable for many reasons,
but here we find an entirely new reason:  because they
facility tunneling and hence forestall saturation with $V_g$.

Figure~\ref{FigIvg}(c) also shows results for a narrow CNT,
$d$=0.8 nm, corresponding to a Schottky contact.
Then there is no step in current, because the current from propagating states
is negligible compared to the tunneling current.
There is only a smooth increase with $V_g$,
with no very clear threshold for turn-on \cite{footnote_vth}.

In summary we have shown that the Ohmic-Schottky crossover in CNT/metal contact occurs with a smooth variation in on-state current, but a sharp change in the transfer characteristics. Typical high-performance devices operate close to the crossover, suggesting an opportunity for further performance improvements via workfunction engineering.
These phenomena are directly relevant to other devices based on low-dimensional semiconductors, such as MoS2 transistors \cite{Kis}, where contacts also play a limiting role in applications.

\newpage

\makeatletter
\makeatletter \renewcommand{\fnum@figure}
{\figurename~S\thefigure}
\makeatother

\renewcommand{\bibnumfmt}[1]{[S#1]}
\renewcommand{\citenumfont}[1]{\textit{S#1}}

\renewcommand{\figurename}{Figure}
\renewcommand{\theequation}{S\arabic{equation}}

{\bf{Supplemental Material}}

\section{\label{append1} Details of the model}

To simulate the current flow in carbon nanotube field effect transistors (CNTFETs) we solve self-consistent Poisson equation for the electrostatics in the wrap-around gate geometry.
For energies outside the bandgap, the distribution function $f^{r}_i(E,x)$ for right-moving carrier resolved by energy $E$ and band index $i$ obeys:
\begin{eqnarray}
0&=&-\frac{df_{i}^{r}(E,x)}{dx}+\frac{f_{i}^{r}(E,x)-f^{0}(E-E_{F}^{M}(x))}{v_{i}(E,x)\tau_{M}(E,x)}
\nonumber \\
&&+\frac{f_{i}^{r}(E,x)-f^{0}(E-E_{loc}(x))}{v_{i}(E,x)\tau_{scat}(E,x)}
\label{eqdfdt_sup}
\end{eqnarray}
where $v_i(E,x)$ is the band velocity and $f^0(x)=(\exp{(x/k_BT)}+1)^{-1}$ is the Fermi-Dirac distribution, where $T=300$ K throughout the paper.
The second term describes an electrical coupling to the metal, which provides a source of current into and out of the nanotube \cite{DattaS}.
Note, that for left-moving carriers $f^{l}_i(E,x)$, the sign of $v_i(E,x)$ in Eq.~(\ref{eqdfdt_sup}) is opposite.
The source and drain are ideal metals with respective Fermi-levels $E_F^{M}=0$ and $E_F^{M}= W_{s}-W_{d}-eV_d $, where $V_d$ is the applied source-drain bias, and $W_s$ and $W_d$ are the workfunctions of the source and drain electrodes.  In Eq.~(\ref{eqdfdt_sup}) we write $E_F^{M}(x)$ to indicate that the coupling (where nonzero) is to whichever contact is locally in contact.  The third term allows thermalization of the carriers toward an equilibrium distribution within a relaxation-time approximation,
with the local quasi-Fermi level $E_{loc}(x)$ found self-consistently.
The timescales for carrier transfer between CNT and metal is determined by the CNT-metal coupling $\eta_M$, via $\tau_M(E,x)= \hbar \eta^{-1}_M(E,x)$.  The scattering time $\tau_{scat}(E,x)= \hbar \eta^{-1}_{scat}(E,x)$ depends on the scattering rate $\eta_{scat}$.

For the propagating states, Eq.~(\ref{eqdfdt_sup}) provides a complete formulation to find distribution function in the working CNT device.
Where tunneling occurs, it is taken into account by appropriate boundary conditions at the classical turning points, see below. The electrostatic potential $\phi(x)$ defines a charge neutrality point (midgap for semiconducting tubes) $E_{NP}(x)$ according to
\begin{eqnarray}
E_{NP}(x)=W_s-W_{c}-e\phi(x)
\label{eqNP}
\end{eqnarray}
where $W_{c}$ is a (metallic) CNT workfunction, here taken to be 4.5 eV.
We treat the CNT bandstructure within a rigid band approximation. So given the nanotube bandstructure, the local electronic structure is fully specified by an energy shift associated with the local potential: $E(k,x)=\varepsilon_k^i+E_{NP}(x)$, where $k$ is the 1D wavevector along the CNT axis.  Here we use $\varepsilon_k^i=\pm (\Delta_i^2+\hbar^2 v_F^2 k^2)^{1/2}$ which has a single parameter $v_F\approx 10^8$ cm/s. A one dimensional wavevector along the CNT axis is $k$ and the bandgap is $2\Delta_i$, where $\Delta_i=i \times 2\hbar v_F/3d$, where $d$ is a CNT diameter and $i=1,2,4,5, ...$ (an integer which is not a multiple of 3).

We can express energy resolved current density $j$ and total current $I$ in the CNT as
\begin{eqnarray}
j_i(E,x)&=&\frac{4e}{h}\left(f_i^r(E,x)-f_i^l(E,x)\right)
\nonumber \\
I(x)&=&\sum_{i}\int_{-\infty}^{\infty} j_i(E,x) dE
\label{eqcurr_sup}
\end{eqnarray}
which at any $x$ is applicable for energies outside the local bandgap.  For a given energy $E$, if $x_1$ and $x_2$ are the classical turning points where the band velocity vanishes, in the forbidden region between points $x_1$ and $x_2$ the tunneling current density is found from the boundary conditions imposing the current conservation:
\begin{eqnarray}
f^l(E,x_1)&=&f^r(E,x_1)R_1+T_1f^0(E-E_F^M)+T_3f^l(E,x_2)
\nonumber \\
f^r(E,x_2)&=&f^l(E,x_2)R_2+T_2f^0(E-E_F^M)
\nonumber \\
&&+T_3f^r(E,x_1)
\label{eq_boun}
\end{eqnarray}
where $E_F^M$ is the Fermi level inside the metal, reflection probabilities are $R_1=1-T_1-T_3$ and $R_2=1-T_2-T_3$.
We approximate tunneling probabilities through the barrier $T_3=P(x_1,x_2)$, to the metal from the left and from the right of the barrier $T_1$ and $T_2$, correspondingly, as:

\begin{eqnarray}
&&T_1=\int_{x_1}^{x_2}\alpha\kappa_M dx P(x_1,x), T_2=\int_{x_1}^{x_2}\alpha\kappa_M dxP(x,x_2),
\nonumber \\
&&P(x,x')=\exp{\left[-\int_{x}^{x'}dx\left(2\kappa+\alpha\kappa_M\right)\right]}
\label{eq_T3}
\end{eqnarray}
where $\kappa=(\Delta_i^2-(E-E_{NP}(x))^2)^{1/2}/(\hbar v_F)$  and $\kappa_M=(\tau_M v_F)^{-1}$ are inverse tunneling and metal coupling lengths, correspondingly.

Eq.~(\ref{eq_T3}) can be derived by assuming that the probability $P(x,x')$ to find an electron at point $x'$, if $P(x,x)=1$,is described by the differential equation $dP(x,x')=-dx'P(x,x')(2\kappa +\alpha/(\tau_{M}v_F))$, where the first term gives carrier reduction due to the reflection and the second term due to the transfer to the metal. We have tested approximation in Eq.~(\ref{eq_T3})  against exact solutions for the tunneling probability from the transfer matrix method.
For abrupt square well potentials, where the error is expected to be maximum, we find at most a factor of $2$ discrepancy.

In general the value of $\alpha$ in Eq.~(\ref{eq_T3}) depends on the energy of the incoming electron far away from the barrier.
It can be shown analytically that for the square well barrier, coupled to the metal by an imaginary energy $E\rightarrow E+i\eta_M$, and an incident carrier momentum $k_1=(E-E_{NP})/(\hbar v_F)$ the value of $\alpha$, in the limit of small $\eta_M$, is given by:
$\alpha=4\frac{\sqrt{k_1^2-k_y^2}}{k_1-\sqrt{k_y^2-\kappa^2}}$, where $k_y=\Delta_i/(\hbar v_F)$. Since the energy $E-E_{NP}$ outside the barrier depends on distance away from the barrier, it leaves some ambiguities in the value of $\alpha$. Just outside the barrier $k_1=k_y$ and $\alpha=0$, unless $\kappa=0$ when $\alpha=\infty$. In general $k_1$ outside the barrier is of the order of $2k_y$ and we are mainly interested in cases with small $\kappa$, so evanescent modes can propagate long distances under the barrier $\kappa\ll k_y$. Therefore, we approximate $\alpha$ by a constant $\alpha=4\sqrt{3}\approx 7$.

\section{\label{append2} Self-consistent CNT doping}

In the absence of tunneling, Eq.~3 of the main text:
\begin{eqnarray}
I_{se}&=& \frac{4e}{h}k_BT\ln{\left[
1+\exp{\left( \frac{E_{Fs}}{k_BT}   \right)}
\right]}
\label{EqImaxS}
\end{eqnarray}
provides a good estimate for the current. The CNT doping level $E_{Fs}$ due to charge transfer from the metal  is
fully specified by the metal/CNT workfunction difference $\Delta W$ and CNT tube diameter. We use electrostatic distance $d_0=2.5$ \AA \ to calculate metal-CNT capacitance $C_M=2\pi\epsilon_0/\ln{\left(1+2d_0/d\right)}$ and charge carrier density
\begin{eqnarray}
\rho&=&g\int_{-\infty}^{-\Delta_1}\frac{f^{0}(-E-E_{Fs}-\Delta_1)}{\pi\hbar v_F\sqrt{E^2-\Delta^2_1}}EdE
\label{eqFermiS}
\end{eqnarray}
which has to be found self-consistently with Eq.~(\ref{eqNP}), where $e\phi(x)=\rho/C_M$ under the metal. The results of the source exhaustion model  \cite{starvation_GuoS,starvation_WongS} can be readily obtained by using degeneracy $g=4$  in Eq.~(\ref{eqFermiS}), which is applicable in the low bias $\vert V_d\vert\ll E_{Fs}$ or in the diffusive limit, when carriers in opposite directions are at equilibrium.
In ballistic channel and high bias $\vert V_d\vert\gg E_{Fs}$, the  left and right moving carriers are described by Fermi distributions with different Fermi levels. Such that the source starvation effect takes place \cite{starvation_FischettiS}, when hot carriers from the drain have  occupancy of unity and don't contribute to the hole carrier density. This can be accounted for by using degeneracy $g=2$ in Eq.~(\ref{eqFermiS}).  We find that the effective $g$ can be determined from the self-consistent solution and in general $2\le g\le4$.


\begin{thebibliography}{}

\bibitem{Martel01}
R.~Martel, V.~Derycke, C.~Lavoie, J.~Appenzeller,
K.~K.~\ Chan, J.~Tersoff, and Ph.~Avouris,
Phys.~Rev.~Lett.~87, 256805 (2001).


\bibitem{Heinze_Schottky} S. Heinze, J. Tersoff, R. Martel, V. Derycke, J. Appenzeller, P. Avouris, Phy. Rev. Lett. {\bf 89}, 106801 (2002).

\bibitem{Joerg} J. Appenzeller, J. Knoch, V. Derycke, R. Martel, S. Wind, and Ph. Avouris, Phys. Rev. Lett. {\bf 89}, 126801 (2002).

\bibitem{Javey03}
Ali Javey, Jing Guo, Qian Wang, Mark Lundstrom, Hongjie Dai,
Nature 424, 654 (2003).

\bibitem{Leonard_APL} A. W. Cummings and F. L\'{e}onard, Appl. Phys. Lett. {\bf 98}, 263503 (2011).

\bibitem{starvation_Guo} J. Guo, S. Datta, M. Lundstrom, M. Brink, P. McEuen, A. Javey, H. Dai, H. Kim, and M. McIntyre,
IEDM Tech. Dig., 711 (2002).

\bibitem{starvation_Fischetti}
M. V. Fischetti, L. Wangt, B. Yut, C. Sachs, P. M. Asbecki, Y. Taurt, and M. Rodwell,
Proc. IEEE IEDM 109, (2007).

\bibitem{starvation_Wong} L. Wei, D. J. Frank, H.-S. P. Wong,  IEEE Trans. on Elect. Dev. {\bf 58}, 2456 (2011).

\bibitem{Zhihong} Z. Chen, J. Appenzeller, J. Knoch, Y.-M. Lin, and P. Avouris, Nano Lett. {\bf 5}, 1497 (2005).


\bibitem{Datta} S. Datta, Quantum Transport Atom to Transistor (Cambridge Univ. Press, 2005).

\bibitem{suppl} See Supplemental Material for a more complete description of the model.





\bibitem{params-Yan}
We focus on the case of a nearly ballistic channel,
taking $\tau_{scat}v_F\gg L_{ch}=300$ nm,
and gate dielectric $\epsilon$=3.9 as for SiO$_2$.
To best match the electrostatics of the
10 nm back-gate geometry of Ref.~\cite{Zhihong}
within our wrap-gate configuration,
we take oxide thickness $t_{ox}$=20 nm;
see Ref.~\cite{Yan}.
We checked that under the biasing conditions and device geometry
used in the experiment the acoustic and optical phonon scattering \cite{VP2005}
with realistic coupling strengths do not change significantly
the values of the on-state current.
We use $\tau_Mv_F=100$ nm to account for the experimental transfer length \cite{Aaron}
and long metal contacts $L_{con}=600$ nm $\gg \tau_Mv_F$. We include two subbands $i=1,2$.
Temperature is $300$ K throughout the paper.

\bibitem{Khomyakov} P. A. Khomyakov, G. Giovannetti, P. C. Rusu, G. Brooks, J. van den Brink, and P. J. Kelly,
Phys. Rev. B {\bf 79}, 195425 (2009).

\bibitem{Giovannetti} G. Giovannetti, P. A. Khomyakov, G. Brooks, V. M. Karpan, J. van den Brink, and P. J. Kelly,
Phys. Rev. Lett. {\bf 101}, 026803 (2008).



\bibitem{Heinze_scaling} S. Heinze, M. Radosavljevic, J. Tersoff, and Ph. Avouris, Phys. Rev. B {\bf  68}, 235418 (2003).




\bibitem{footnote_vth}
We define $V_t$ in this regime as the value of $V_g$
at which a tangent to the $I$-$V_g$ curve crosses x-axis at a 0.5 V above $V_g$.
In the ohmic regime (linear transfer characteristic) this reduces to the conventional definition.






\bibitem{Yan} R.-H. Yan, A. Ourmazd, and K. F. Lee, IEEE Trans. Electron Dev. {\bf 39}, 1704 (1992).

\bibitem{VP2005}
V.~Perebeinos, J.~Tersoff, and Ph.~Avouris,
Phys.~Rev.~Lett.~94, 086802 (2005).

\bibitem{Aaron} A. D. Franklin and Z. Chen, Nature Nano, {\bf 5}, 858 (2010).


\bibitem{Leonard_pinning} F.~L\'{e}onard and J.~Tersoff, Phys.~Rev.~Lett.~84, 4693 (2000).

\bibitem{bokor} Y.-C. Tseng, K. Phoa, D. Carlton, and J. Bokor, Nano Lett. {\bf 6}, 1364 (2006).


\bibitem{Klimeck} G. Fiori, G. Iannaccone, and G. Klimeck IEEE Transactions on Electron Devices, {\bf 53}, 1782 (2006).


\bibitem{footnote_alpha} In general the value of $\alpha$ depends on the energy of the incoming electron far away from the barrier.
It can be shown analytically that for the square well barrier, coupled to the metal by an imaginary energy $E\rightarrow E+i\eta_M$, and an incident carrier momentum $k_1=(E-E_{NP})/(\hbar v_F)$ the value of $\alpha$, in the limit of small $\eta_M$, is given by:
$\alpha=4\frac{\sqrt{k_1^2-k_y^2}}{k_1-\sqrt{k_y^2-\kappa^2}}$, where $k_y=\Delta_i/(\hbar v_F)$. Since the energy $E-E_{NP}$ outside the barrier depends on distance away from the barrier it leaves some ambiguities in the value of $\alpha$. Just outside the barrier $k_1=k_y$ and $\alpha=0$, unless $\kappa=0$ when $\alpha=\infty$. In general $k_1$ outside the barrier is of the order of $2k_y$ and we are mainly interested in cases with small $\kappa$, so evanescent modes can propagate long distances under the barrier $\kappa\ll k_y$. Therefore, we approximate $\alpha$ by a constant $\alpha=4\sqrt{3}\approx 7$.

\bibitem{Kis}  B. Radisavljevic, A. Radenovic, J. Brivio, V. Giacometti, and A. Kis, Nature Nano {\bf 6}, 147 (2011).


\end{thebibliography}

\begin{thebibliography}{}

\bibitem{DattaS} S. Datta, Quantum Transport Atom to Transistor (Cambridge Univ. Press, 2005).

\bibitem{starvation_GuoS} J. Guo, S. Datta, M. Lundstrom, M. Brink, P. McEuen, A. Javey, H. Dai, H. Kim, and M. McIntyre,
IEDM Tech. Dig., 711 (2002).

\bibitem{starvation_WongS} L. Wei, D. J. Frank, H.-S. P. Wong,  IEEE Trans. on Elect. Dev. {\bf 58}, 2456 (2011).

\bibitem{starvation_FischettiS}
M. V. Fischetti, L. Wangt, B. Yut, C. Sachs, P. M. Asbecki, Y. Taurt, and M. Rodwell,
Proc. IEEE IEDM 109, (2007).

\end{thebibliography}
\end{document}